\documentclass[floats,floatfix,showpacs,amssymb,prd,twocolumn,superscriptaddress,nofootinbib,nolongbibliography,reprint]{revtex4-1}

\usepackage{amssymb,amsmath,verbatim,mathtools,needspace,enumitem,etoolbox,graphicx,physics,microtype,afterpage,xspace,tabularx,lmodern,multirow}
\usepackage{gensymb}
\usepackage[normalem]{ulem}
\usepackage[dvipsnames, usenames]{xcolor}
\definecolor{linkcolor}{rgb}{0.0,0.3,0.5}
\usepackage[unicode, colorlinks=true, linkcolor=linkcolor, citecolor=linkcolor, filecolor=linkcolor, urlcolor=linkcolor, linktocpage, breaklinks]{hyperref}
\usepackage[all]{hypcap}
\usepackage{subfigure}
\usepackage[T1]{fontenc}
\usepackage[utf8]{inputenc}
\usepackage[usenames,dvipsnames]{xcolor}
\hypersetup{colorlinks=true,citecolor=romared,linkcolor=romared,urlcolor=romared}

\definecolor{romared}{RGB}{142,0,28}

\newcommand{\be}{\begin{equation}}
\newcommand{\ee}{\end{equation}}

\def\be{\begin{equation}}
\def\ee{\end{equation}}
\newcommand{\beq}{\begin{eqnarray}}
\newcommand{\eeq}{\end{eqnarray}}

\usepackage{aas_macros}
\usepackage{makecell}
\usepackage{soul}
\usepackage{amssymb}

\usepackage{lipsum}

\newcommand{\quotes}[1]{{``#1''}}

\newcommand{\software}[1]{\texttt{#1}}
\newcolumntype{Y}{>{\centering\arraybackslash}X}
\newcommand{\tabcolsepcustom}{4pt}

\begin{document}
\title{The impact of relativistic corrections on the detectability of dark-matter spikes with gravitational waves}

\begin{abstract}
  Black holes located within a dark matter cloud can create overdensity regions known as dark matter spikes. The presence of spikes modifies the gravitational-wave signals from binary systems through changes in the gravitational potential or dynamical friction effects.  We assess the importance of including relativistic effects in both the dark matter distribution and the dynamical friction.  As a first step we numerically calculate the particle dark matter spike distribution in full general relativity, using both Hernquist and Navarro-Frenk-White profiles in a Schwarzschild background, and we produce analytical fits to the spike profiles for a large range of scale parameters.  Then we use a post-Newtonian prescription for the gravitational-wave dephasing to estimate the effect of relativistic corrections to the spike profile and to the dynamical friction. Finally we include the torques generated by dynamical friction in fast-to-generate relativistic models for circular extreme mass-ratio inspirals around a nonspinning black hole. We find that both types of relativistic corrections positively impact the detectability of dark matter effects, leading to higher dephasings and mismatches between gravitational-wave signals with and without dark matter spikes.
\end{abstract}

\author{Nicholas Speeney}
\email{nspeene1@jhu.edu}
\affiliation{Department of Physics and Astronomy, Johns Hopkins University, 3400 N. Charles Street, Baltimore, Maryland, 21218, USA}

\author{Andrea Antonelli}
\email{aantone3@jhu.edu}
\affiliation{Department of Physics and Astronomy, Johns Hopkins University, 3400 N. Charles Street, Baltimore, Maryland, 21218, USA}

\author{Vishal Baibhav}
\email{vishal.baibhav@northwestern.edu}
\affiliation{Center for Interdisciplinary Exploration and Research in Astrophysics (CIERA)
and
Department of Physics and Astronomy, 
Northwestern University, 
1800 Sherman Ave, 
Evanston, IL 60201
USA}

\author{Emanuele Berti}
\email{berti@jhu.edu}
\affiliation{Department of Physics and Astronomy, Johns Hopkins University, 3400 N. Charles Street, Baltimore, Maryland, 21218, USA}

\date{\today}
\maketitle

\section{Introduction}\label{sec:intro}

It has been long established that galaxies contain distributions of dark matter (DM) which affect their formation history~\cite{Blumenthal:1984bp,2008gady.book.....B}. Models for the distribution of DM ranging from numerical fits to $N$-body cosmological simulations suggest that the mass distribution is peaked near the galactic center and tapers off as some power of $1/r$, where $r$ is the distance from the halo's center. Such density distributions can be generically described by
\begin{equation}
\label{initial density}
\rho (r) = \rho_0(r/a)^{-\gamma} (1+(r/a)^\alpha)^{(\gamma - \beta)/\alpha},
\end{equation}
where $\alpha, \beta$ and $\gamma$ are parameters that control the small-$r$, large-$r$, and transition-region  behavior of the density distribution respectively, while
 
$a$ and $\rho_0$ are scale factors. The specific choice  $(\alpha,\beta,\gamma)=(1,4,1)$ corresponds to the Hernquist 
profile~\cite{Hernquist:1990be}, whereas $(\alpha,\beta,\gamma)=(1,3,1)$ corresponds to the Navarro-Frenk-White (NFW) profile~\cite{Navarro:1996gj}.

It is interesting to consider how the  DM density changes in the presence of a black hole (BH), as is most relevant in the neighborhood of the galactic center. In an early study,

Gondolo and Silk proposed a Newtonian scheme for the redistribution of cold DM around the central BH starting from a power-law distribution, with some relativistic ansatz for the allowed energies and angular momenta of the  DM~\cite{Gondolo:1999ef}.

They found that the  DM forms a cusp (\quotes{spike}) structure in the presence of the black hole, with a density peak roughly near $4 R_s$ (with $R_s$ the Schwarzschild radius) and a steep cutoff at $4 R_s$, below which the  DM particles fall into the BH. Sadeghian et al.~\cite{Sadeghian:2013laa} later performed a fully general relativistic calculation for a Hernquist profile and selected values of the scale parameters, finding some important qualitative differences in the spike features compared to Ref.~\cite{Gondolo:1999ef}. In the relativistic treatment the spikes can get significantly closer to the central black hole, showing a density peak close to $2 R_s$ rather than $4R_s$, and an even greater overdensity near the central BH.

These results bear important consequences for phenomena near the galactic cores, such as binaries formed by a compact object (of mass $m$) orbiting around the more massive BH (of mass $M_\text{BH}$) in the galactic center. The mass ratio $q\equiv m/M_\text{BH}$ of these binaries is typical of intermediate mass-ratio inspirals (IMRIs)~\cite{Amaro-Seoane:2007osp}, roughly composed of $q\sim \mathcal{O}(10^{-3}-10^{-4})$ systems, and extreme mass-ratio inspirals (EMRIs), $q\sim \mathcal{O}(10^{-5}-10^{-6})$~\cite{Babak:2017tow}.
The former are binaries that can be observed with upcoming next-generation detectors on Earth (e.g., the Einstein Telescope~\cite{Punturo:2010zz} and Cosmic Explorer~\cite{Reitze:2019iox}) and in space (LISA~\cite{Audley:2017drz}), while the latter can be observed exclusively with LISA.
The effect on these sources from the presence of a DM overdensity is twofold: it modifies the gravitational well \quotes{felt} by the secondary compact object, and it induces dynamical friction (DF) drag forces that tend to slow down the binary's trajectory~\cite{Chandrasekhar:1943ys,Barausse:2014tra}.

It has been estimated that the change in trajectory for IMRIs in the LISA band due to torques generated by such a drag force could be used to detect and measure the effect of DM with gravitational waves (GWs)~\cite{Eda:2014kra,Kavanagh:2020cfn,Coogan:2021uqv}. While these analyses are explicitly carried out in the case of IMRIs, for which we expect higher spike densities~\cite{Eda:2013gg,Eda:2014kra}, EMRIs could also be used to constrain DM models~\cite{Hannuksela:2019vip}.

The analyses of Refs.~\cite{Kavanagh:2020cfn,Coogan:2021uqv} make some approximations regarding the signals' evolutions, modelling them within Newtonian gravity. Both IMRIs and EMRIs require, however, fully relativistic waveform models that are capable of tracking their phase accurately~\cite{Barack:2009ux,Barack:2018yvs}. Work is currently underway to provide the community with such accurate models~\cite{Pound:2019lzj,Miller:2020bft,Warburton:2021kwk,Wardell:2021fyy}. The \texttt{FastEMRIWaveform} (FEW) models are a promising framework for fast EMRI/IMRI waveform generation~\cite{Chua:2018woh,Chua:2020stf,Katz:2021yft}. They are currently available for eccentric adiabatic orbits in a Schwarzschild background in the Black Hole Perturbation Toolkit~\cite{BHPToolkit} (see the documentation in~\cite{few, michael_l_katz_2020_4005001} for more details). These codes are modular and engineered to include any additional physical effects (due either to the binary dynamics or its environment) as they become available. 

Along with including relativistic corrections to signal trajectories in vacuum, one must also pay attention to relativistic corrections in the modelling of environmental effects. Extending upon the work of Ref.~\cite{Sadeghian:2013laa}, it is important to create a catalog of relativistic DM spikes that does not depend on a specific choice of the scale parameters. Moreover, it has been pointed out in recent numerical simulations~\cite{Traykova:2021dua} and analytical work~\cite{Vicente:2022ivh} that relativistic corrections to the DF force, usually employed in the form found by Chandrasekhar in the Newtonian limit~\cite{Chandrasekhar:1943ys}, may also play an important role for the accurate modelling of environmental effects. Relativistic expressions for the DF have been considered in Refs.~\cite{Barausse:2007ph,Hui:2016ltb}.

The goal of this work is to include relativistic corrections to the DM-spike models and the DF force within an already existing relativistic waveform generation scheme for EMRIs/IMRIs, while assessing their relative importance in describing DM-dressed signals. More in detail:
\begin{itemize}
    \item We create a catalog of relativistic DM-spikes, and use them to produce fits valid for a wide range of scale parameters for both initially Hernquist and NFW profiles (Sec.~\ref{sec:DM_spikes}).
    \item We include the DF torques induced by DM in post-Newtonian (PN) GW waveforms for binaries in circular orbits within a PN treatment, and in the relativistic trajectories of the FEW models (Sec.~\ref{sec:torques}).
    \item We assess the relative importance of different relativistic DM-spike models and corrections to Chandrasekhar's DF formula within a large region of the parameter space. For this, we use estimates of the GW cycles and, for a restricted set of realistically-detectable EMRIs, we compute dephasings and mismatches that account for LISA noise (Sec.~\ref{sec:results}). 
\end{itemize}
We conclude in Sec.~\ref{sec:conclusions} that relativistic corrections to both the DM spike model and DF have an appreciable impact, and that DM models can be efficiently added to the already available, fast-to-generate EMRI/IMRI models of Refs.~\cite{few,michael_l_katz_2020_4005001}.

\section{Dark Matter Spike Calculation and Scaling Laws}
\label{sec:DM_spikes}

In this section, we will briefly review the formalism used to calculate the  DM density in the presence of a BH. The Newtonian calculation can be found in Ref.~\cite{Gondolo:1999ef}, but we are primarily interested in the fully relativistic calculation of the spikes set forth by Sadeghian et al.~\cite{Sadeghian:2013laa}.
Starting from a density profile for the DM given by Eq.~\eqref{initial density}, we calculate the phase space distribution function of this density profile, either analytically if possible (as is the case for an initial Hernquist profile), or numerically using Eddington inversion if there is no analytical solution (as is the case for an initially NFW profile). The key equation to note for the Eddington inversion method is~\cite{2008gady.book.....B,Lacroix:2018qqh}

\begin{align}
\label{Eddington inversion}
f(\mathcal{E}) &=\frac{1}{\sqrt{8}\pi^2}\left ( \frac{1}{\sqrt{\mathcal{E}}} \left [ \frac{\text{d}\rho}{\text{d} U}\right ]_{U=0} + \int^{\mathcal{E}}_0  \frac{\text{d}U}{\sqrt{\mathcal{E}-U}} \frac{\text{d}^2\rho}{\text{d}U^2}\right ) \nonumber \\
&= \frac{1}{\sqrt{8}\pi^2}\left ( \frac{\text{d}}{\text{d}\mathcal{E}} \int^{\mathcal{E}}_0  \frac{\text{d}U}{\sqrt{\mathcal{E}-U}} \frac{\text{d}\rho}{\text{d}U}\right),
\end{align}
where $f(\mathcal{E})$ is the phase space distribution function, $\mathcal{E}$ is the energy, and $U$ is the gravitational potential associated with the initial DM distribution.

The central black hole grows adiabatically in the cloud of DM if the orbital timescale of a DM particle is much greater than the time it takes for the black hole to grow in the center of the cloud. Because the process is adiabatic, we can compute the radial, azimuthal, and polar adiabatic invariants $I=\{I_r,I_\theta,I_\phi\}$ of the system before and after the black hole grows. Before growth, the DM halo is all that matters for calculating the invariants. The relevant invariants for the halo are~\cite{Sadeghian:2013laa} 

\begin{equation}
\label{halo invariants}
\begin{aligned}
I^{\text{H}}_r(E',L') & = \oint dr \sqrt{2E'-2U-{L'}^2/r^2},\\
I^{\text{H}}_{\theta}(L',L'_z) &=  \oint d\theta \sqrt{{L'}^2-{L'}_z^2 \sin^{-2}{\theta}}=2\pi(L'-L'_z),\\
I^{\text{H}}_{\phi}(L'_z) &= \oint d\phi L'_z = 2\pi L'_z,
\end{aligned}
\end{equation}
where $E', L', L'_z$ are the energy, total angular momentum, and angular momentum along the $z$-direction, respectively.

After growth, the background geometry is assumed to be purely Schwarzschild, and the fully relativistic analysis leads us to the following expressions for the adiabatic invariants of the BH:
\begin{align}
\label{BH invariants}
I^{\text{BH}}_r(E,L) & = \oint dr \sqrt{E^2-(1-2G M_{\text{BH}} /r)(1+L^2/r^2)},\nonumber\\
I^{\text{BH}}_{\theta}(L,L_z) &= 2\pi(L-L_z),\nonumber\\
I^{\text{BH}}_{\phi}(L_z) &= 2\pi L_z.
\end{align}
 
Equating the $\theta$ and $\phi$ components before and after the growth tells us that the angular momenta of the DM particles will be the same throughout, $L'=L, L_z'=L_z$, as is expected for growth on a Schwarzschild background and collisionless DM. The energies however will change, being related by the equation 
\begin{equation}
\label{eq:radial_action_equality}
I_r^{\text{H}}(E',L) = I_r^{\text{BH}}(E,L). 
\end{equation} 

Using Eq.~\eqref{eq:radial_action_equality}, we can find $E'$ as a function of $E$. This allows us to calculate the final halo density profile by integrating over the phase space distribution via~\cite{Sadeghian:2013laa}
\begin{align}
\label{relativistic final density}
& \rho(r) = \frac{1}{\sqrt{2}(2\pi)^2}\left ( \frac{M_{\text{DM}}}{a^3}\right ) \frac{a}{r-R_S} \\ & \times \int_0^{\tilde{\epsilon}_{\text{max}}}d\tilde{\epsilon}[1-(GM_{\text{DM}}/a)\tilde{\epsilon}] \int_{\tilde{L}_{\text{min}}^2}^{\tilde{L}_{\text{max}}^2}{d\tilde{L}^2 \frac{\tilde{f}_\text{H}(\tilde{\epsilon}'(\tilde{\epsilon},\tilde{L}))}{\sqrt{\tilde{L}^2_{\text{max}}-\tilde{L}^2}}},  \nonumber
\end{align}
where $M_\text{DM}$ is the total initial halo mass, $M_\text{BH}$ is the mass of the black hole, $\tilde{\epsilon}$, $\tilde{L}$, and $ \tilde{f}_\text H$ are the rescaled energy, angular momentum, and halo distribution function, respectively (exact definitions can be found in Ref.~\cite{Sadeghian:2013laa}), and $a$ is the scale factor appearing in the initial density of Eq.~\eqref{initial density}. A similar expression for the Newtonian version of this calculation can be found in Eq.~(4.16) of Ref.~\cite{Sadeghian:2013laa}.

\begin{figure}[t]
    \centering
    \includegraphics[width=.49\textwidth]{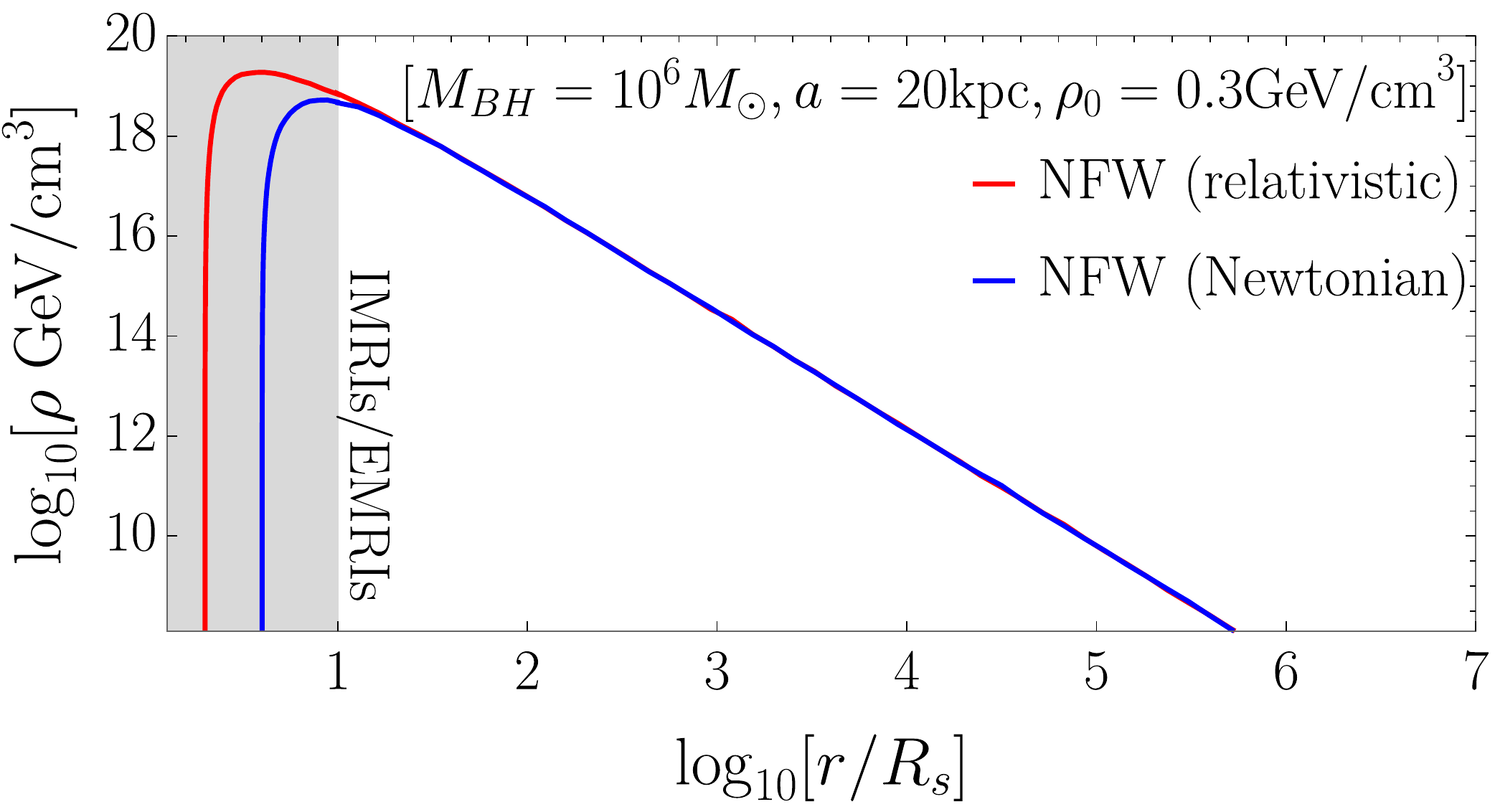}
    \includegraphics[width=.49\textwidth]{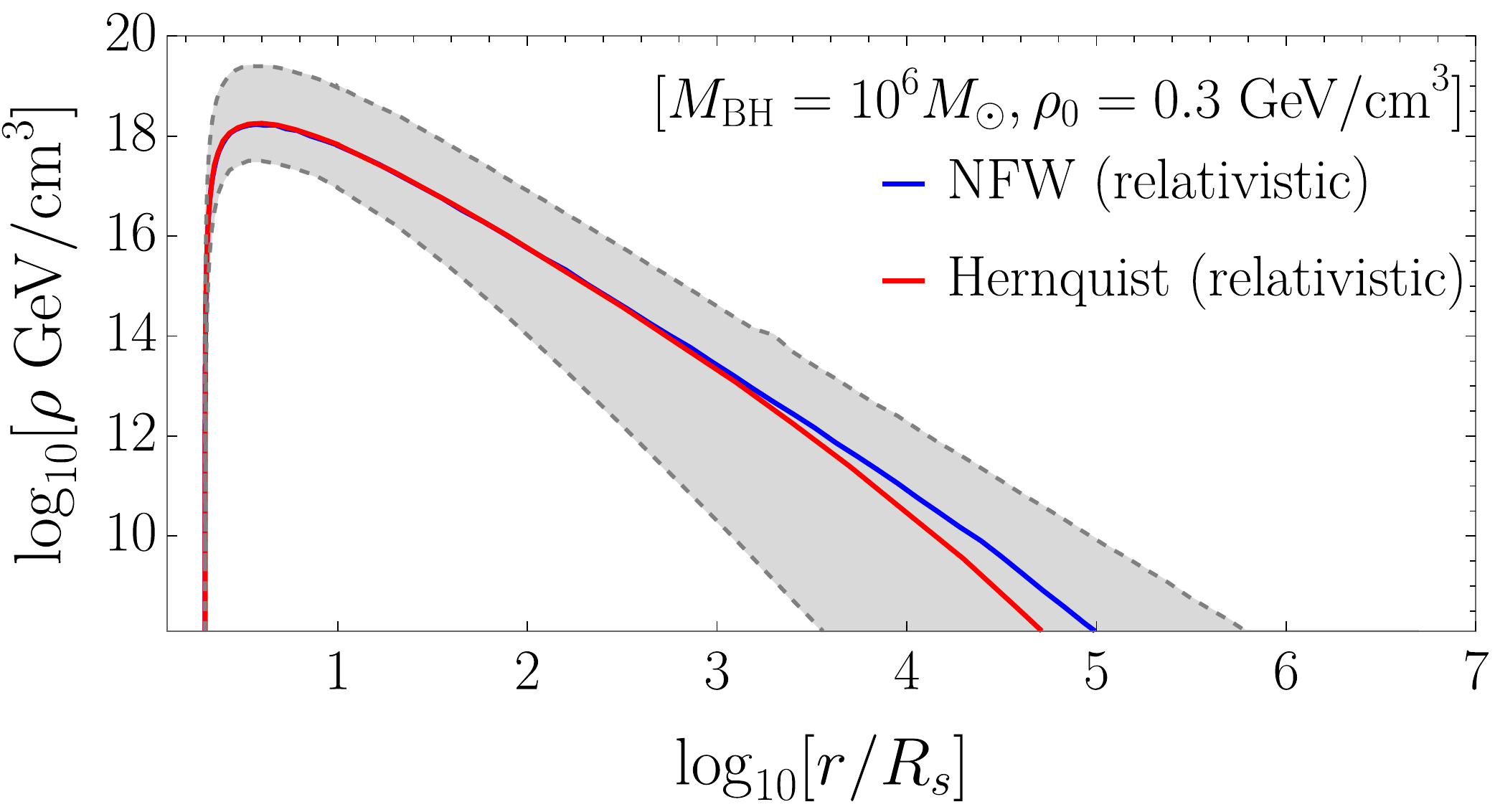}
        \caption{DM spikes for initially Hernquist and NFW distributions. Top panel: the relativistic spike description is important for IMRIs/EMRIs at radii $r \lesssim \mathcal{O}(10)R_s$ (shaded area). Note that $R_s = 9.608 \times 10^{-11} (M_{\text{BH}}/10^6M_\odot) \text{kpc}$, meaning that we are always in the $r \ll a$ regime for the masses and orbital radii $r\lesssim 10^2 R_s$ relevant for LISA. 
        Bottom panel: the NFW and Hernquist spike models are indistinguishable for radii $r\ll a$, but as one approaches the $r \gg a$ regime the spike slopes change to those of the initial distribution.  The red and blue curves are generated by setting $a=0.01 \text{kpc}$. The shaded region in the bottom panel is bounded from below (above) by the Hernquist curve generated using $a=0.002\text{kpc}$ ($a=50\text{kpc}$), to demonstrate the effect of changing the value of $a$. Note that the lower bound corresponds to an extremely small value of $a$. More realistic choices in the range $0.005\text{kpc} < a < 50\text{kpc}$ do not yield such an extreme slope for $r\lesssim 10^2 R_s$.}
\label{DM spikes}
\end{figure}

Figure~\ref{DM spikes} shows the results from calculating the final density distribution of the DM as described above for initial Hernquist and NFW profiles. The curves shown are generated by setting $\rho_0= 0.3 \text{GeV/cm}^3$ and $M_{\text{BH}} = 10^{6} \text{M}_\odot$.
 
The top panel is calculated for a nominal value of $a=20$kpc and an NFW profile, and is meant to show that the spikes do not fall off as sharply as in the Newtonian treatment of Ref.~\cite{Gondolo:1999ef}. We have checked that the same applies to the case of Hernquist profiles, reproducing results in Ref.~\cite{Sadeghian:2013laa}. The impact of relativistic corrections to the spikes is investigated further in later sections of the paper.
In the bottom panel, we fix $a=0.01$kpc to highlight some qualitative features between relativistic NFW and Hernquist spikes that appear as $r\sim a$ (roughly corresponding to $\log_{10}[r/R_s]\sim 3.5$).
There, the DM density falls off more sharply for the Hernquist profile. 
For both the Hernquist and NFW spikes, the slope in the far region approaches the slope of the pre-growth distribution's far region, which is to be expected since the central black hole's influence becomes very small in this regime. 
 
While the most relevant portion of the distribution for GW astrophysics applications is the inner region ($r\ll a$), the long-range spike behavior is of relevance for cosmology or galaxy scale astrophysics, and is thus interesting to point out.
 
Using the catalogs of curves we generated for various parameter choices, we find effective scaling laws for the inner regions of the DM spikes, valid when $r \lesssim 50M_{\text{BH}}$ and $0.01\text{kpc} \leq a$. The scaling laws are found by separately varying $\rho_0, ~a, ~\text{and} ~M_{\text{BH}}$, and fitting the height change of Hernquist and NFW curves as a function of the varied parameter. We have checked that the scaling still works even when all parameters are varied at once, by directly comparing to a curve generated numerically using Eq.~\eqref{relativistic final density}. The effective scaling is given by
\begin{equation}
\begin{aligned}
\label{spike_scalings}
\rho = \bar{\rho}\, 10^\delta \left( \frac{\rho_0}{0.3 \text{GeV/cm}^3}\right)^{\alpha} \left( \frac{M_{\text{BH}}}{10^6 M_{\odot}}\right)^{\beta} \left( \frac{a}{20 \text{kpc}}\right)^{\gamma},
\end{aligned}
\end{equation}
 with parameters $\alpha = 0.335 (0.331)$, $\beta = -1.67 (-1.66)$, $\gamma = 0.31 (0.32)$, $\delta = -0.025 (-0.000282)$ for Hernquist (NFW) spike profiles. The overall scale is found by fitting a reference curve generated with ($\rho_0,M_{\text{BH}},a$)=($0.3\text{GeV}/\text{cm}^3,10^6M_\odot,20\text{kpc}$) using \texttt{Mathematica}'s built-in \texttt{NonlinearModelFit} function with three parameters $A,~w,~q$, and is given by 

\begin{equation}
\label{spike_parameterization}
\begin{aligned}
\bar{\rho} = A \left(1-\frac{4\eta}{\tilde{x}} \right)^w\left(\frac{4.17\times 10^{11}}{\tilde{x}}\right)^q,
\end{aligned}
\end{equation}

with $\eta =1 (2)$ for relativistic (Newtonian) DM profiles and $\tilde{x}\equiv r/M_{\text{BH}}$.
 
The amplitude $A$ and exponents $w$, $q$ are given in Table~\ref{tab:fit_parameters}. 

\begin{table}
\label{tab:fit_parameters}
\caption{Fitting parameters for the NFW and Hernquist DM density spike profiles, valid when $r \ll a$ and $0.01\text{kpc} \leq a$.}
\begin{tabularx}{\linewidth}{ @{\hspace{\tabcolsepcustom}}c | @{\hspace{\tabcolsepcustom}}c  @{\hspace{\tabcolsepcustom}}c  |@{\hspace{\tabcolsepcustom}}c @{\hspace{\tabcolsepcustom}}c}
\hline \hline  
\makecell{Fit \\ Parameter}& \makecell{Hernquist \\ (Newton)} & \makecell{Hernquist \\ (rel.)} &\makecell{NFW \\ (Newton)} &\makecell{ NFW \\ (rel.)} \\ \hline \hline
$A$ ($10^{-43}M_{\odot}^{-2}$) & $4.87$ & $7.90$  & $1.60$ & $6.42$ \\ \hline
$w$ & $2.22$ & $1.83$ & $2.18$ & $1.82$ \\ \hline
$q$ & $1.93$ & $1.90$ & $1.98$ & $1.91$ \\ \hline
\end{tabularx}
\end{table}

\section{Binary Inspiral within the DM Spike}
\label{sec:torques}

Now that we can generate density profiles for a large portion of the parameter space of initial halo and black hole parameters through Eq.~\eqref{spike_scalings}, we move on to calculating the effect of these spikes on a binary inspiral. In Sec.~\ref{sec:SPA}, we primarily follow the prescription laid out in Refs.~\cite{Eda:2013gg,Eda:2014kra} to obtain a measure of the dephasing cycles which accounts for the changes in both the gravitational potential and the DF induced by the DM spike. In Sec.~\ref{sec:FEW_traj} we briefly discuss how the FEW models can be changed to accommodate environmental effects.
We restrict our attention to the circular-orbit limit, mainly for simplicity. This limitation is not unphysical, as Ref.~\cite{Becker:2021ivq} has shown that DF circularizes inspiraling orbits within a DM spike (see however how DF could enhance the eccentricity of binaries within the cusp in Ref.~\cite{Yue:2019ozq}).

\subsection{Estimating the number of cycles from DM-spike effects}
\label{sec:SPA}

We start by calculating the radial equation of motion, using the Newtonian expressions for circular binaries, as
\begin{align}
\label{eq:radial_EoM}
\ddot{r} = -\frac{\text{d} \Phi}{\text{dr}} + \frac{l^2}{r^3} &= -\frac{G M_{\text{BH}}}{r^2} \left [1 + \epsilon q(r)\right ] + r \omega_s^2, \\
\epsilon q(r) &\equiv M_{\text{DM}}(r)/M_{\text{BH}},
\end{align}
where $\omega_s$ is the orbital frequency, $l$ the angular momentum, and $M_{\text{DM}}(r)$ is the mass of the DM spike profile enclosed within a sphere of radius $r$. Note that overdots denote differentiation with respect to time throughout the paper.
One can check that $M^{\text{total}}_{\text{DM}} \ll M_{\text{BH}}$ for the regions of parameter space we consider, so that the quantity $\epsilon q(r)\lesssim 10^{-7}$ throughout the range of the spike in Fig.~\ref{DM spikes}.

Rearranging \eqref{eq:radial_EoM} and expanding appropriately in $\epsilon$, we  find an expression for $r(\omega_s)$:
\begin{equation}
\label{r_of_omega}
r(\omega_s) = r_0 \left ( 1 + \frac{1}{3} \epsilon q(r_0)  + \mathcal{O}(\epsilon^2) \right ),
\end{equation} 
where we have defined $r_0 = (G M_{\text{BH}}/\omega_s^2)^{1/3}$. Note that throughout this work, we limit our calculations to a first-order expansion in $\epsilon$. We next consider the evolution of the $r$-coordinate for circular orbits evolving adiabatically during the inspiral, so that $\dot{r} \neq 0$. The energy of the orbit changes as the radius of the orbit decreases, following the energy balance equation
\begin{equation}
\label{energy_balance}
\dot{E}_{\text{orbit}} = \dot{E}_{\text{GW}} +\dot{E}_{\text{DF}}.
\end{equation}

The orbital energy loss is given by the standard Newtonian expression for a circular orbit, which using the definitions \eqref{eq:radial_EoM}-\eqref{r_of_omega} can be rewritten as 
\begin{equation}
\label{orbital_energy_loss}
\dot{E}_{\text{orbit}} = \frac{G M_{\text{BH}}}{r^2}  \frac{\mu \dot{r}}{2} \left [ 1+\epsilon q(r) + \epsilon r q'(r) + \mathcal{O}(\epsilon^2) \right ].
\end{equation}

Here, $\mu = m M_{\text{BH}}/(m + M_{\text{BH}})$ is the reduced mass and a prime refers to a derivative with respect to $r$. The GW energy loss formula is given at lowest PN order by the quadrupole formula, which for circular orbits reads
\begin{equation}
\label{GW_energy_loss}
\dot{E}_{\text{GW}} = -\frac{32}{5} G \frac{\mu^2}{c^5} r^4 \omega_s^6,
\end{equation}
where $\omega_s$ is the orbital frequency of the source.
In later sections we consider the effect of higher-order PN corrections~\citep{Blanchet:2013haa}.  
The energy loss due to DF comes from the inspiraling compact object interacting with the DM. The DM particles create a wake behind the massive object that decelerates it.
This energy loss can be thought of as a drag force on the orbiting object, such that~\cite{Chandrasekhar:1943ys}
\begin{equation}
\label{DF_drag_force}
\dot{E}_{\text{DF}}=\mathbf{v} \cdot \mathbf{F}_{\text{drag}}= -4\pi \frac{G^2 \mu^2 \rho(r)}{v} \xi(v) \ln{\Lambda},
\end{equation}
where $\ln{\Lambda}$ is the Coulomb logarithm, defined by $\ln{\Lambda} \approx b_{\text{max}} v_{\text{typ}}^2/(G \mu) \approx 3$~\cite{1943ApJ....97..255C}. Here, $b_{\text{max}}$ is the maximum impact parameter and $v_{\text{typ}}$ is the typical velocity of the stellar-mass object. The factor $\xi(v)$ is a relativistic correction to the DF, taken to be $\xi(v) = \gamma^2(1+\zeta v^2/c^2)^2$, where $\gamma =\sqrt{1-\zeta v^2/c^2}$ (see, e.g.,~\cite{Barausse:2007ph,Traykova:2021dua}). This factor accounts for the increase in deflection angle of the DM (treated as a fluid) as the orbiting body moves through the DM, and also accounts for the relativistic momentum of the fluid as seen by the orbiting mass. $\zeta$ is chosen to be 1 (0) when turning on (off) the relativistic corrections. We omit here a pressure-correcting term to the DF introduced in Ref.~\cite{Traykova:2021dua}, as Ref.~\cite{Vicente:2022ivh} argues that this term is unnecessary when treating DF in a self-consistent relativistic manner.

The DF effect can be compared to the GW effect by computing the relative strength

\begin{equation}
\label{DF_correction_to_energy_loss}
\frac{\dot E_{\text{DF}}}{\dot E_{\text{GW}}} = \frac{c_{\text{DF}}}{c_{\text{GW}}} \frac{(1+G M_{\text{BH}}/r_0c^2)^2}{(1-G M_{\text{BH}}/r_0c^2)}r_0^{11/2} \rho(r), 
\end{equation}
with
\begin{equation}
 c_{\text{GW}} = \frac{64}{5} \frac{G^3}{c^5} \mu  M_{\text{BH}}^2,
\qquad c_{\text{DF}} = 8 \pi  \mu  \sqrt{\frac{G}{M_{\text{BH}}^3}} \ln{\Lambda}.
\end{equation}
The coefficients $c_{\text{GW}}$ and $c_{\text{DF}}$ encode the GW and DF effects, respectively. 
In Sec.~\ref{sec:FEW_traj}, we use Eq.~\eqref{DF_correction_to_energy_loss} to compute torques in the FEW model codes by treating the DF as a correction to the energy loss.

Using Eq.~\eqref{r_of_omega}, we invert Eq.~\eqref{energy_balance} to get the frequency evolution equation
\begin{equation}
\label{fsdot}
\dot{f_s} = \frac{3 \sqrt{G M_{\text{BH}}}}{4 \pi} \left[\dot{f_s}^{(0)}(r_0) + \epsilon \dot{f_s}^{(1)}(r_0)+ \mathcal{O}(\epsilon^2)\right], 
\end{equation}
 where we have defined
 \begin{align}
 \label{f_s_definitions}
 \dot{f_s}^{(0)}&= \frac{c_{\text{GW}}}{r_0^{11/2}} + c_{\text{DF}} \frac{(2r_0+\zeta R_s)^2}{2r_0(2r_0-\zeta R_s)}\rho, \\
 \dot{f_s}^{(1)}&=  \frac{2 c_{\text{GW}}}{3 r_0^{11/2}} \left[ q - 2r_0 q' \right ]- c_{\text{DF}} \frac{(2r_0+\zeta R_s)}{6r_0 (2r_0-\zeta R_s)}  \Xi(\rho),\nonumber
 \end{align}
 with
 \begin{align}
     \Xi(\rho)&=r_0(\zeta^2R_s^2-4r_0^2)(4\rho q'-q\rho')\\
     &+ q \rho (\zeta^2 R_s^2 + 12r_0 \zeta R_s -12r_0^2) \nonumber
 \end{align}
 and $r_0$ defined as before, except that now we replace $\omega_s$ with $f_s = \omega_s/2\pi$. Notice that $f_s = f/2$ is the source orbital frequency, and $f$ is the GW frequency. Also, the $q$'s and $\rho$'s along with their derivatives are functions of $r_0$, which we have omitted in these expressions. 
We can split the various contributions to Eq.~\eqref{f_s_definitions} as coming from GR (the first term of $\dot{f_s}^{(0)}$), DF (the terms proportional to $c_\text{DF}$), and contributions from the spike to the gravitational potential (the first term of $\dot{f_s}^{(1)}$).
By ignoring the DF and DM contributions (i.e., by setting $c_{\text{DF}}=\epsilon=0$) we recover the usual quadrupole formula for GW frequency evolution~\cite{2009GReGr..41.1667H}.
One could use Eq.~\eqref{fsdot} to find the full waveform and phase in the stationary phase approximation. The quantity of interest for us is the orbital phase $\phi_{s}(f)$, related to the number of GW cycles $\mathcal{N}_\text{cycles}(f)$ by
\begin{equation}
\label{eq:Ncycles_def}
\mathcal{N}_\text{cycles}(f) = 
\frac{\phi_s(f)}{\pi}=\int_f^{f_\text{ISCO}}\frac{f'}{\dot{f}'}df'. 
\end{equation}
In what follows, we use this expression to calculate the number of cycles for the variety of models presented in Table~\ref{tab:fit_parameters}. We will often split the contributions from DF and the change in the gravitational potential by turning off the appropriate contributions to the chirping frequency \eqref{f_s_definitions} in the integration of Eq.~\eqref{eq:Ncycles_def}. We compare these to estimates for the cycles obtained from well-known GR corrections to Eq.~\eqref{f_s_definitions} at 1PN and 2PN orders~\cite{Blanchet:1995ez,Berti:2004bd}.

\subsection{Adiabatic trajectories with EMRI models}
\label{sec:FEW_traj}

While the waveform model developed in the previous section is a good guide for the detectability of DM effects, it degrades in accuracy as we reach the strong-field regime due to fundamental limitations of the underlying PN expansions~\cite{Yunes:2008tw,Zhang:2011vha,Sago:2016xsp}. Since we are mostly interested in understanding the relativistic corrections in this region of spacetime, we also employ relativistic models for EMRI waveforms based on the Teukolsky equation~\cite{Teukolsky:1973ha,Press:1973zz}. Specifically we use the \texttt{FastEMRIWaveform} (FEW) packages~\cite{Chua:2018woh,Chua:2020stf,Katz:2021yft,few}. 
These models are more accurate than PN waveforms in the strong-field regime, as they track both the trajectory of the inspiral and the waveform generation, and they have been built to be fast-to-generate and modular enough to include either additional physical information about the signal or environmental effects. In this work, we add the effect of DF without compromising the overall speed of generation of the FEW signals. In this sense, the models used here are good proxies for what could actually be used in LISA data analysis.

In what follows, we perform a detectability analysis with a handful of selected EMRI configurations.
Our reference trajectory for a signal in vacuum is obtained through the \texttt{SchwarzEccFlux} model, while we keep the waveform generation formalism quadrupolar. The trajectory is described by a set of 17 parameters, $\vec\theta=\{M_\text{BH},m, a_1, a_2, p_0, e_0, \iota, D_L, \theta_S, \phi_S, \theta_K, \phi_K, \Phi_{\phi}^{0}, \Phi_{\theta}^{0}, \Phi_{r}^{0}\}$ \cite{Katz:2021yft}. We let the primary and secondary masses $M_\text{BH}$ and $m$, initial separation $p_0$, and distance $D_L$ vary for the different EMRI configurations we consider. We  {assume} the binary to be nonspinning and in a quasicircular orbit by setting the primary spin $a_1$, secondary  {spin} $a_2$ and eccentricity $e_0$ to zero. We do not include any inclination from the equatorial plane, i.e. our inclination angle is always $\iota = 0$. We set the polar and azimuthal sky location angles to a nominal value $\theta_S = \phi_S = \pi/4$, while  {the spin-orientation angles} $\theta_K$ and  $\phi_K$ vanish in our setting. Finally we initialize the polar, azimuthal and radial phases to $\Phi_{\phi}^{0}= \Phi_{\theta}^{0}= \Phi_{r}^{0}=0$. The FEW codes then solve for the time evolution of orbital separation $p(t)$ and phases $\Phi_{\phi}(t), \Phi_{\theta}(t), \Phi_{r}(t)$ over the observation time period $T_\text{obs}$~\cite{Katz:2021yft}. 

We are interested in estimating the changes in phases between a signal in vacuum, as described above, and a signal with the same configuration in which the secondary compact object moves within the DM spike. The DF in fact changes the trajectory of the secondary, inducing a torque that modifies the angular momentum and energy fluxes. We modify the FEW codes to account for this torque, changing the  {energy} fluxes to
\begin{equation}
     {
    \dot E_{\text{GW}} \rightarrow \left(1+ \frac{\dot E_{\text{DF}}}{\dot E_{\text{GW}}}\right)\dot E_{\text{GW}},
    }
\end{equation}
where we use Eq.~\eqref{DF_correction_to_energy_loss}. This modification is valid only as an estimate for the circular orbits we are employing. More careful modelling is required when including noncircular-orbit features.
The two extra parameters we must include in the codes are the scale parameters $\rho_0$ and $a$. All in all, we consider  {8} models: Hernquist and NFW profiles, in each class including relativistic effects in both DF force and spike, only in the spike,  {only in the DF,} or in neither.  
This model choice allows us to explore when relativistic corrections are important to detect DM effects, which we do in the following section.

\section{Results}
\label{sec:results}

\begin{figure}
    \centering
    \includegraphics[width=\linewidth]{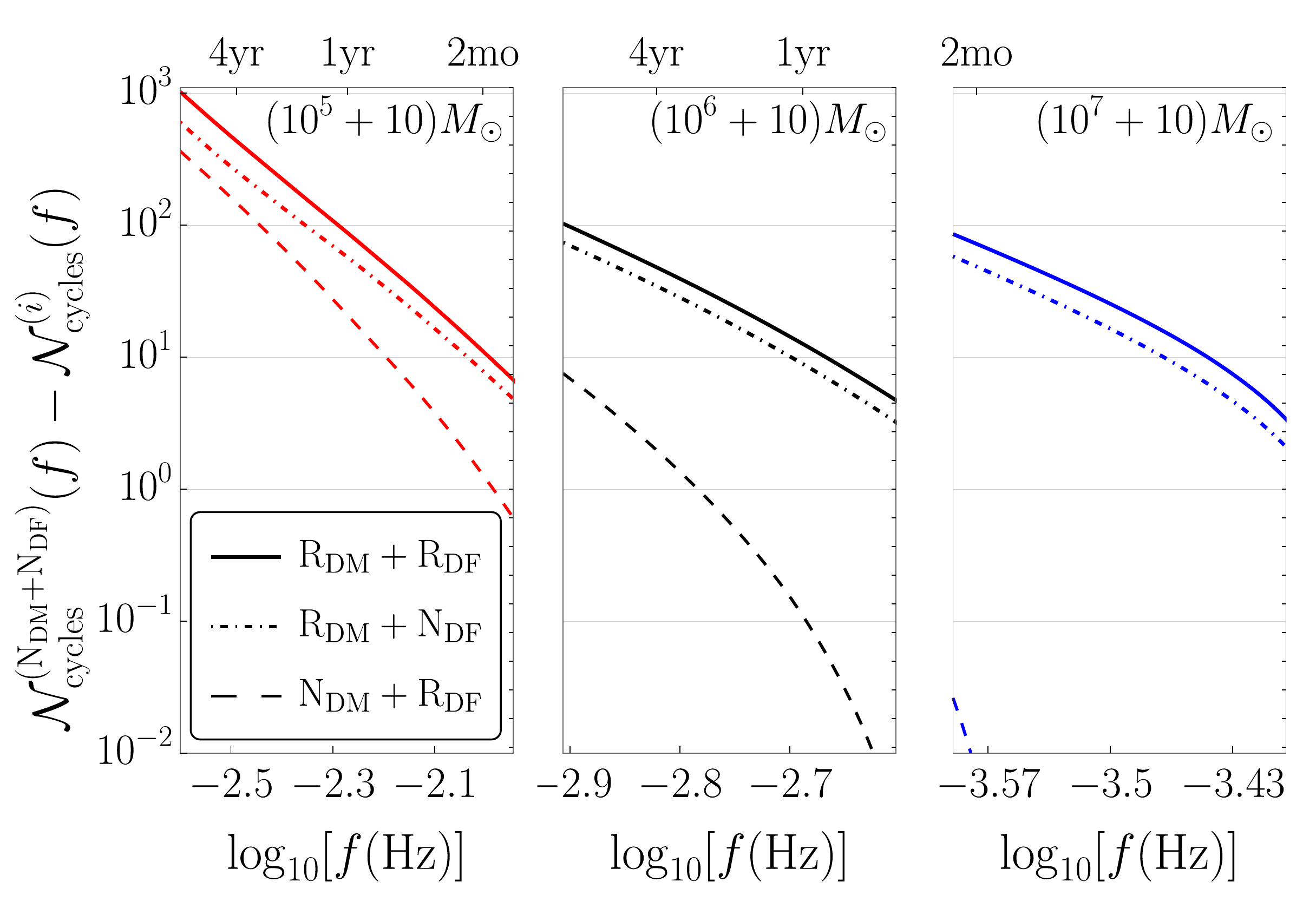}
    \qquad
\caption{Variation in the GW cycles due to including relativistic and Newtonian prescriptions for the DM and DF, compared to the fully Newtonian description, as a function of the GW frequency (bottom axis) and time to merger (top axis). The relativistic correction to the DM spike induces larger dephasings than the relativistic correction to the DF, however both effects are important for $M_\text{BH}\lesssim 10^7 M_\odot$. For $M_\text{BH}=10^7 M_\odot$ the relativistic contribution to the DF is negligible, and barely visible in the plot.}
    \label{fig:cycles}
\end{figure}

\begin{figure}[h]
\includegraphics[width=\linewidth]{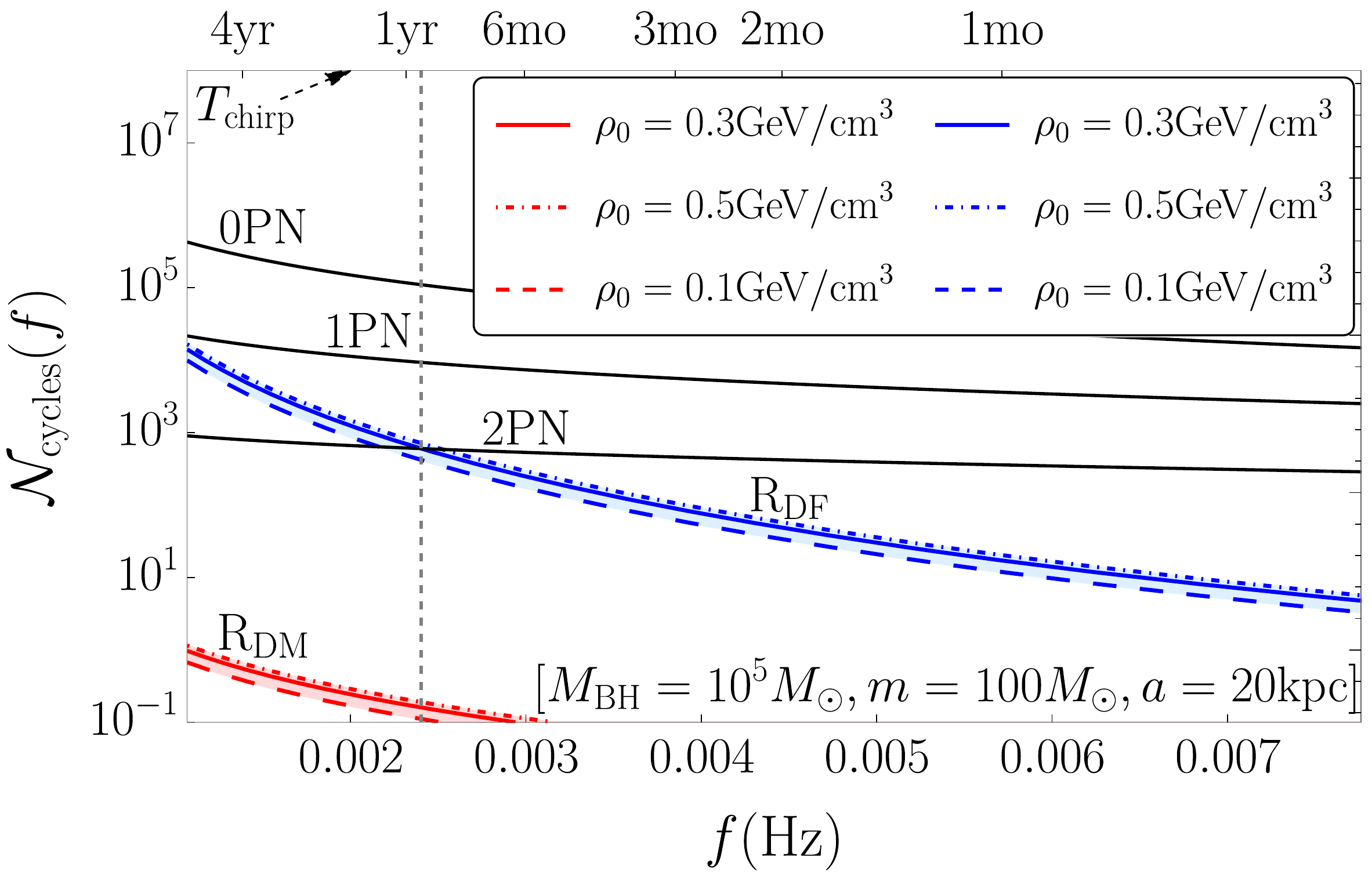}
\caption{Various contributions to $\mathcal{N}_\text{cycles}$ for a $(10^5+100)M_{\odot}$ binary system, using both a relativistic DM spike and relativistic DF. The lower $x$-axis denotes the frequency at which we start observing the system, and assumes that we track that system all the way to the ISCO. The upper $x$-axis is the time to merger. The effect of DM purely on the gravitational potential (the curves labelled $\text{R}_{\text{DM}}$) is far less important than the DF contribution (the curves labelled $\text{R}_{\text{DF}}$). All DM-induced effects are below the 2PN contribution, unless we are able to observe the system for $\gtrsim 1\text{yr}$, at which point the DF becomes dominant over 2PN contributions. As we increase the observing time (moving to the left on the plot), the DF effect becomes more and more important.}
\label{fig:dephasing}
\end{figure}

\begin{table}
\setlength{\tabcolsep}{0.5pt}
 \caption{The value of $\rho_0$ needed to produce $1$ cycle of dephasing for multiple observation times and BH masses. As the observation time increases, smaller values of $\rho_0$ are sufficient to have observable effects. Likewise, smaller BH masses correspond to a larger number of observable GW cycles (cf. Table~\ref{tab:table} below): secular effects due to the DM are enhanced, and again smaller values of $\rho_0$ can lead to observable effects.}
\begin{tabularx}{\linewidth}{ @{\hspace{\tabcolsepcustom}}c  @{\hspace{\tabcolsepcustom}}c  @{\hspace{\tabcolsepcustom}}c  @{\hspace{\tabcolsepcustom}}c}
\hline
\hline
  $\rho_0 ~[\text{GeV}/\text{cm}^3]$& \makecell{$(10^7+50)M_\odot$}& \makecell{$(10^6+50)M_\odot$} & \makecell{$(10^5+50)M_\odot$}  \\ \hline
$1$ wk & $7249$ & $179$ & $0.091$ \\ 
$1$ mo & $103$ & $1.8$ & $0.00014$ \\ 
$3$ mo & $4.9$ & $0.037$ & $9.4\times 10^{-7}$ \\ 
$6$ mo & $0.75$ & $0.0025$ & $3.9\times 10^{-8}$ \\ 
$1$ yr & $0.11$ & $0.00015$ & $4.6\times 10^{-9}$ \\ \hline \hline
\end{tabularx}
\label{tab:crit_rho0}
\end{table}

\begin{table*}
\setlength{\tabcolsep}{1.235em}
 \caption{Number of GW cycles $\mathcal{N}_\text{cycles}$  {contributed by selected} PN orders in vacuum, compared to DM and DF contributions. The rows labelled \quotes{R$_{\text{DF}}$} and \quotes{N$_{\text{DF}}$} are the DF contributions in the presence of a relativistic DM spike generated using $a$=$20$kpc, with the BH mass listed in each column. We report estimates for different values of $\rho_0$. The frequency values correspond to a 4-year observation time, and in all cases, we consider a $50 M_\odot$ secondary. Round brackets denote corrections to the cycles which are dominant over 2PN terms in vacuum general relativity, but subdominant with respect to 1PN terms. 
 We have not included values of $M_\text{BH}< 10^5M_\odot$ since they would require a prescription for the halo feedback~\cite{Kavanagh:2020cfn}, which we do not include in this work.}
\begin{tabularx}{\linewidth}{ c | c c c | c c c | c c c }
\hline \hline  
$M_\text{BH}$& \multicolumn{3}{c|}{\makecell{$10^7M_\odot$}}& \multicolumn{3}{c|}{\makecell{$10^6M_\odot$}} & \multicolumn{3}{c }{\makecell{$10^5M_\odot$}}  \\ \hline \hline
$f_{\text{in}}$ [Hz]&\multicolumn{3}{c|}{$0.00024$} & \multicolumn{3}{c|}{$0.00093$} & \multicolumn{3}{c}{$0.00179$} \\ 
$f_{\text{fin}}$ [Hz]&\multicolumn{3}{c|}{$0.00044$} & \multicolumn{3}{c|}{$0.00438$} & \multicolumn{3}{c}{$0.04383$} \\ \hline 
0 PN &\multicolumn{3}{c|}{$282132$}& \multicolumn{3}{c|}{$214542$} & \multicolumn{3}{c}{$360269$} \\ 
1 PN &\multicolumn{3}{c|}{$83782$}& \multicolumn{3}{c|}{$39972$} & \multicolumn{3}{c}{$25329$} \\ 
1.5 PN &\multicolumn{3}{c|}{$-175262$}& \multicolumn{3}{c|}{$-67858$} & \multicolumn{3}{c}{$-27886$} \\ 
2 PN &\multicolumn{3}{c|}{$15502$}& \multicolumn{3}{c|}{$4960$} & \multicolumn{3}{c}{$1399$} \\ \hline 
$\rho_0$ [GeV/cm$^3$]& \multicolumn{1}{c|}{$0.1$}& \multicolumn{1}{c|}{$0.3$}& \multicolumn{1}{c|}{$0.5$}& \multicolumn{1}{c|}{$0.1$}& \multicolumn{1}{c|}{$0.3$}& \multicolumn{1}{c|}{$0.5$}& \multicolumn{1}{c|}{$0.1$}& \multicolumn{1}{c|}{$0.3$}& \multicolumn{1}{c}{$0.5$}
\\ \hline 
R$_\text{DM}$&$0.15$&$0.21$&$0.25$ & $0.17$&$0.24$ &$0.28$&$0.54$ &$0.79$&$0.93$ \\ 
N$_\text{DM}$&$0.04$ &$0.05$&$0.06$ &$0.05$ & $0.08$ &$0.09$ &$0.53$ & $0.77$& $0.91$ \\ 
R$_\text{DF}$&$-14.7$ &$-21.2$&$-25.1$ &$-67$ &$-97$&$-115$ &($-3266$) &($-4682$)&($-5531$) \\
N$_\text{DF}$&$-10.1$ &$-14.6$&$-17.3$ &$-54$ &$-79$&$-94$ &($-3047$) &($-4370$)&($-5163$) \\ \hline
\end{tabularx}
\label{tab:table}
\end{table*}

\begin{figure}[h]
\includegraphics[width=\linewidth]{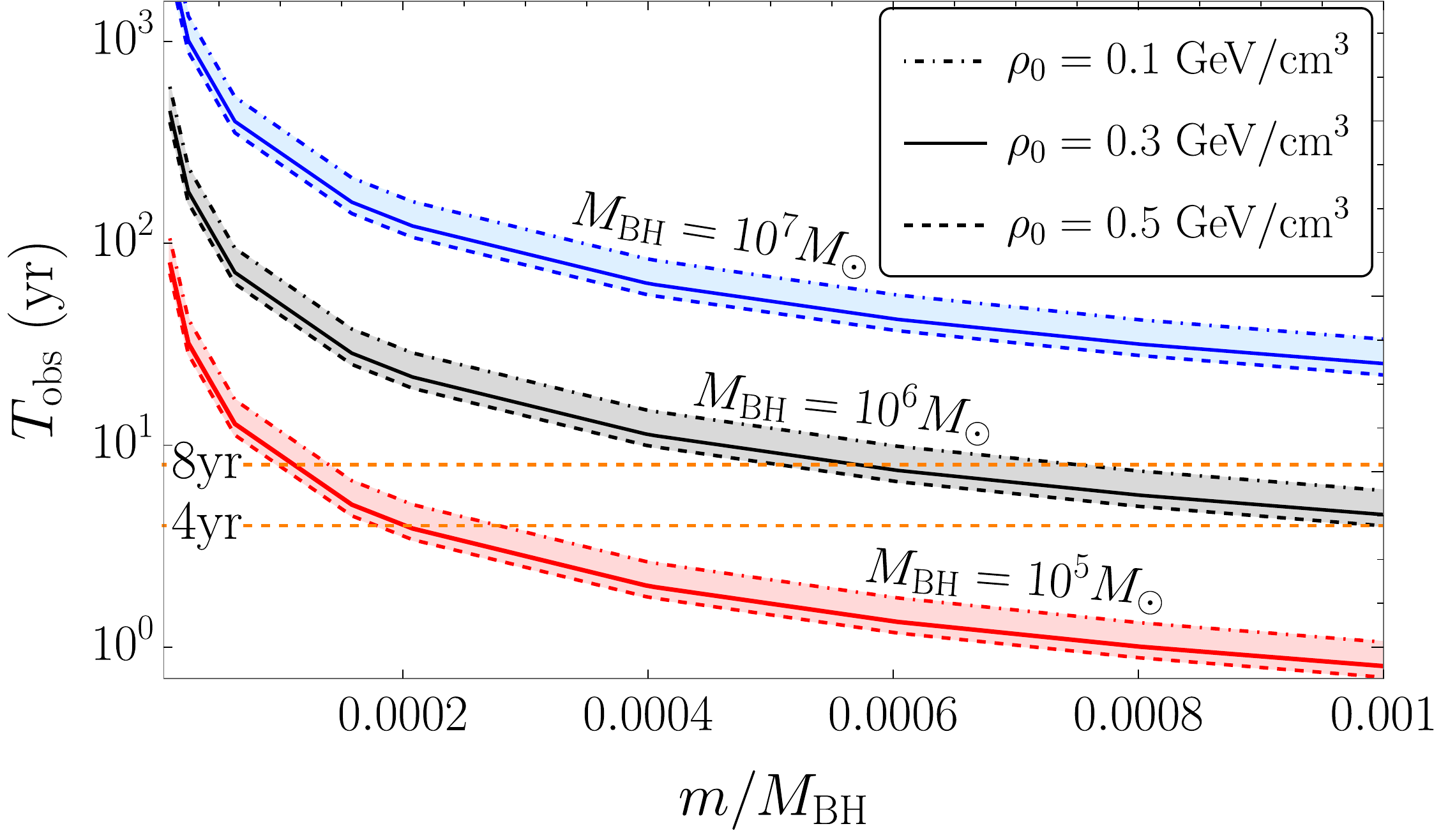}
\caption{Necessary observation time such that the DM+DF effects become dominant over the 2PN contributions to the dephasing cycles. The nominal (4-year) observation time for the LISA mission, as well as a more optimistic 8-year observation time, are reported as dashed horizontal lines. The 2PN crossings can occur for  $(10^5 + >10)M_\odot$ systems for realistic LISA observing timescales.}
\label{fig:PN_comp_to_DM}
\end{figure}

\begin{figure}[th!]
\centering
\includegraphics[width=\linewidth]{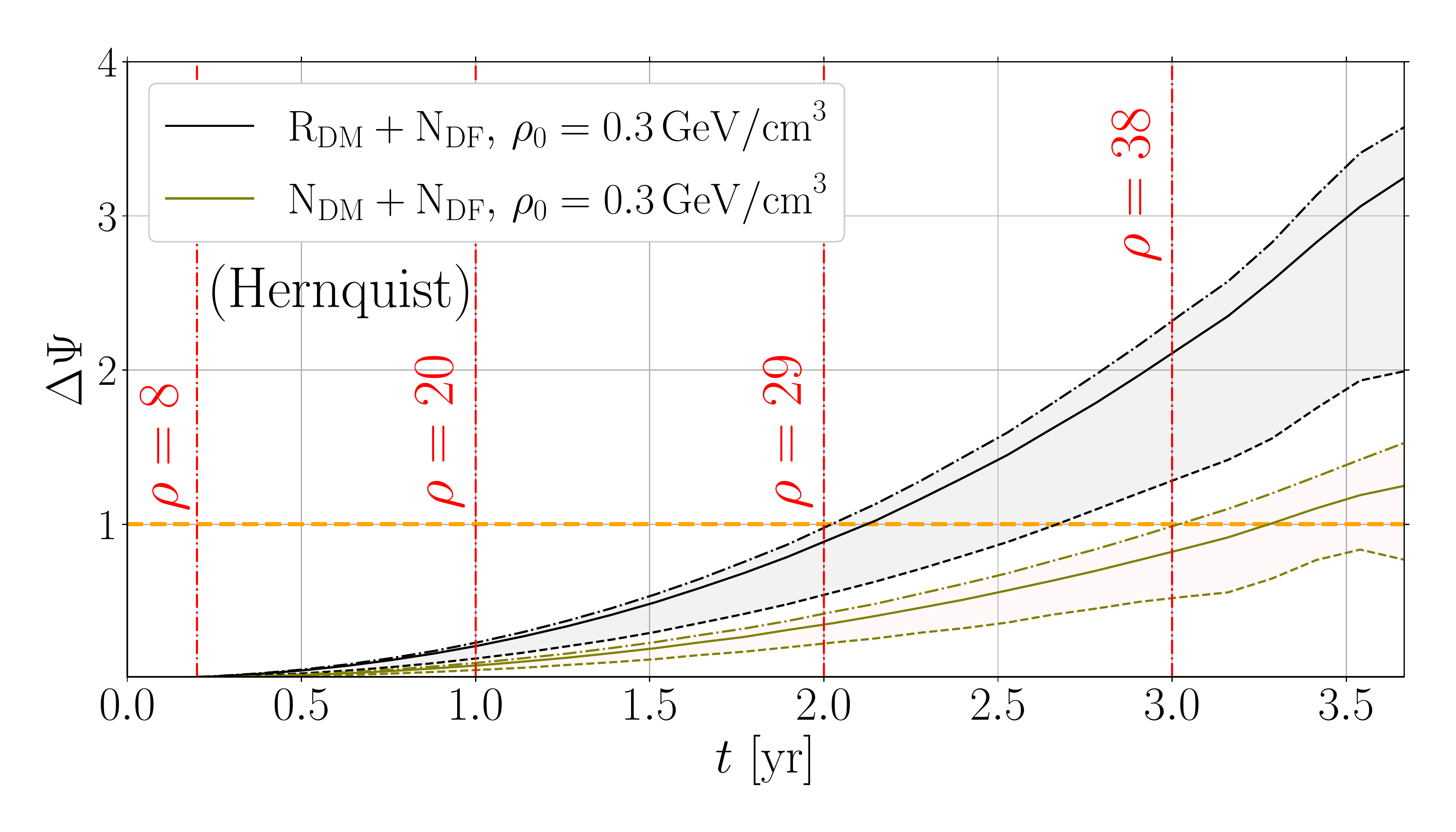}
\includegraphics[width=\linewidth]{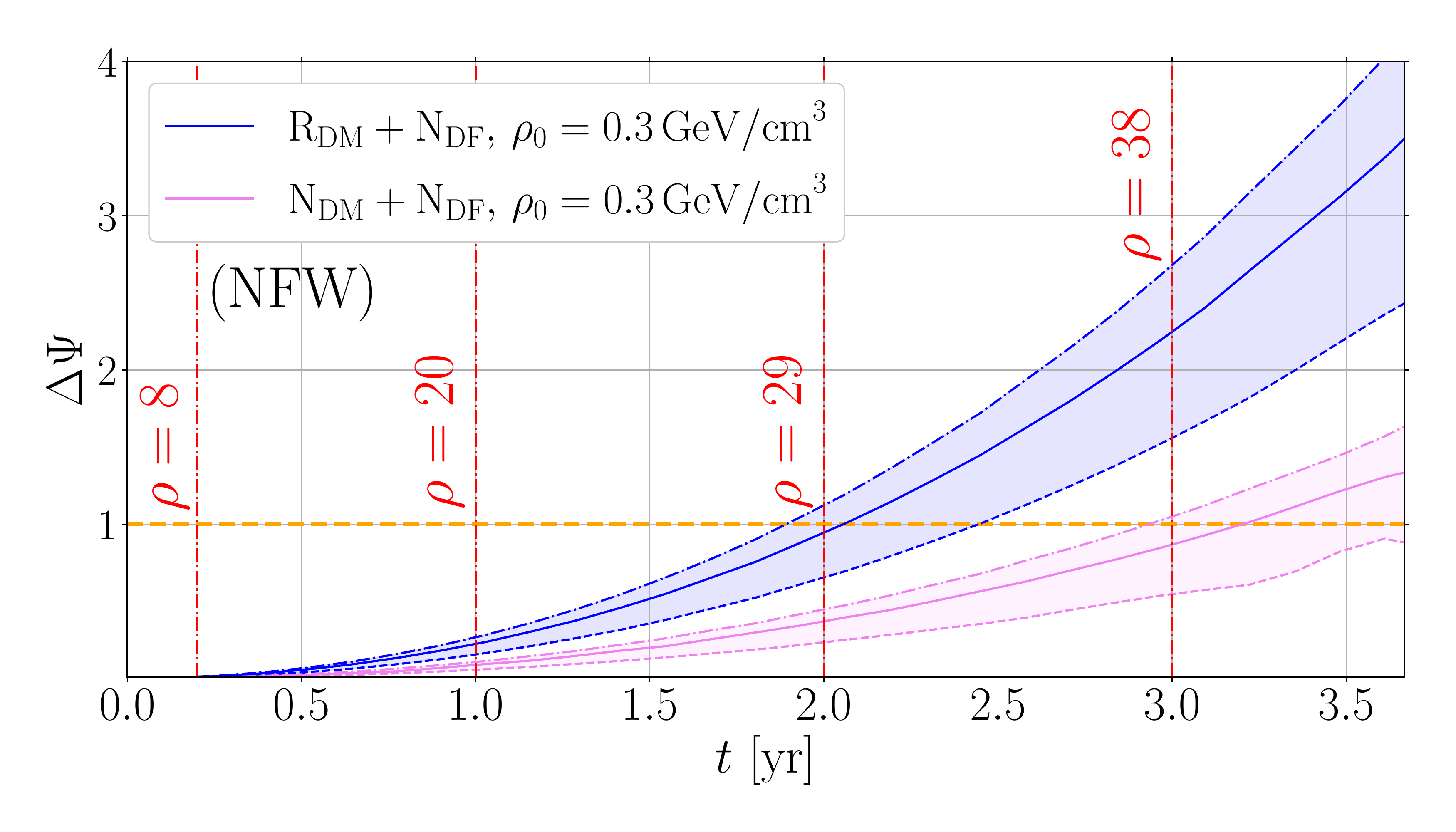}
\includegraphics[width=\linewidth]{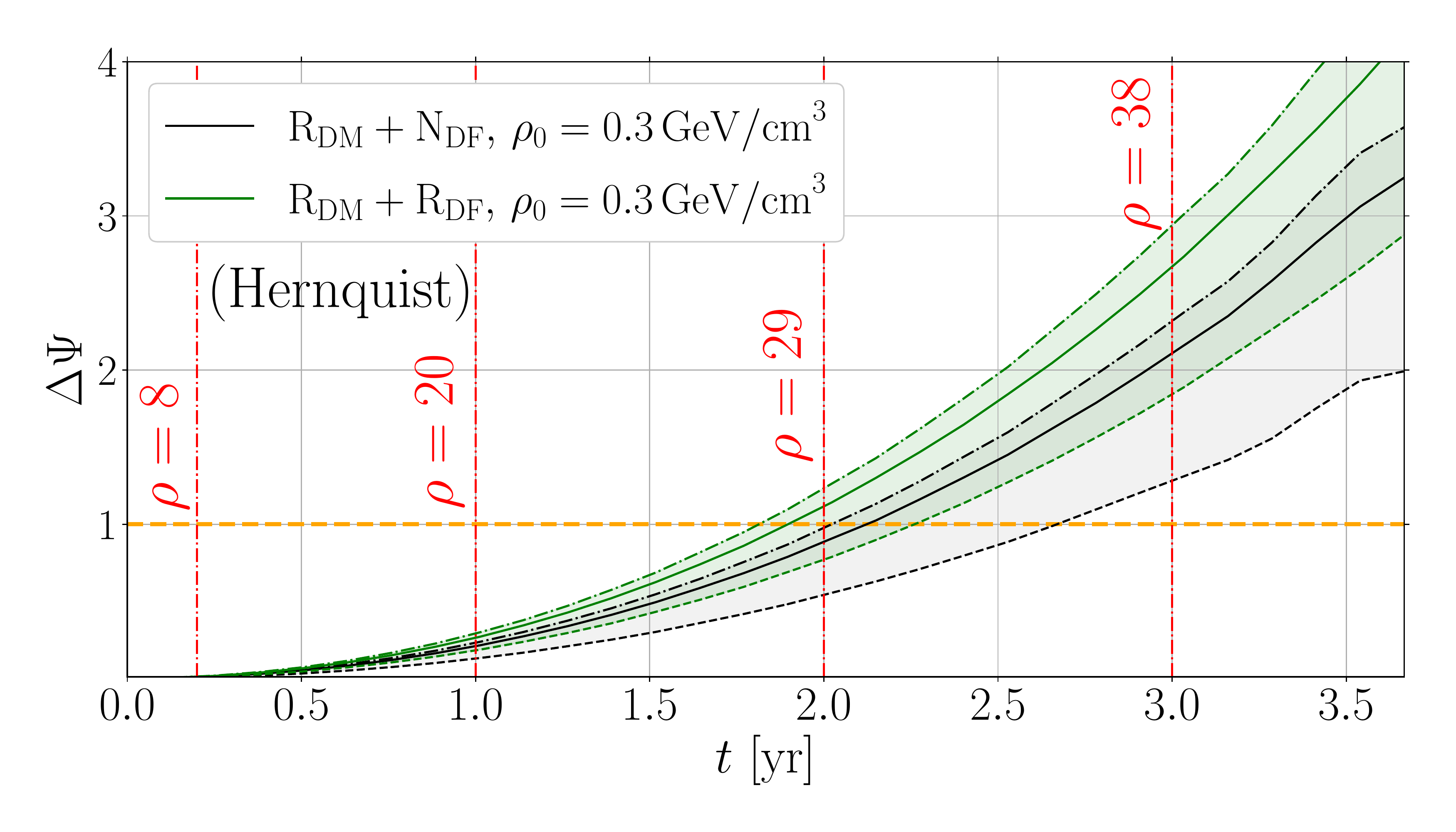}
\caption{
Top: dephasings using both relativistic and Newtonian Hernquist profiles, with a Newtonian (Chandrasekhar-like) DF force.
Middle: same for an NFW profile. 
Bottom: dephasings from relativistic Hernquist profiles, with and without relativistic corrections to the DF. 
Dephasing effects are (roughly) detectable above the horizontal dashed line, which corresponds to $\Delta\Psi = 1$. We use $a=20$~kpc, $M_{\rm BH}=10^6 M_\odot$, $m=10M_\odot$, and a distance $D_L=1$~Gpc in all panels. The solid lines are the dephasing for the reference value $\rho_0 = 0.3$ GeV/cm$^3$, while the shaded area is bounded from above (below) by dephasings with values $\rho_0=$0.5 (0.1) GeV/cm$^3$.
The vertical lines correspond to the accumulation of the signal-to-noise ratio (SNR) throughout the observation.
}
\label{fig:EMRI_dephasing_Hernquist}
\end{figure}

\begin{figure*}[th!]
\centering
\includegraphics[width=\linewidth]{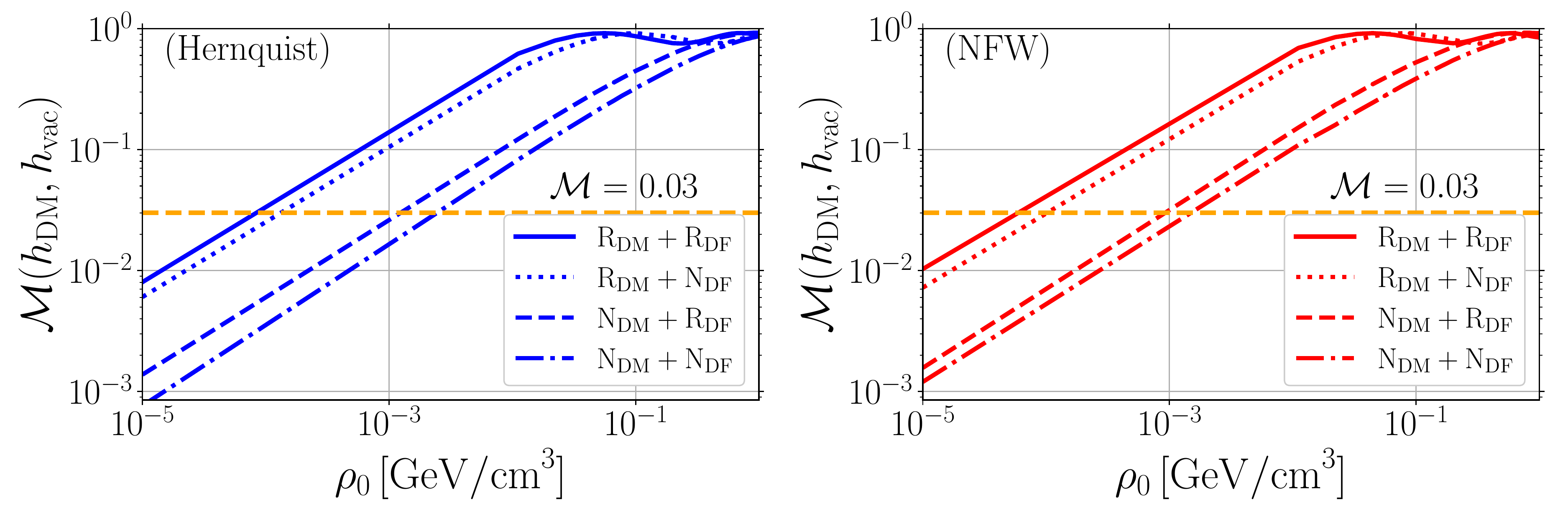}
\includegraphics[width=\linewidth]{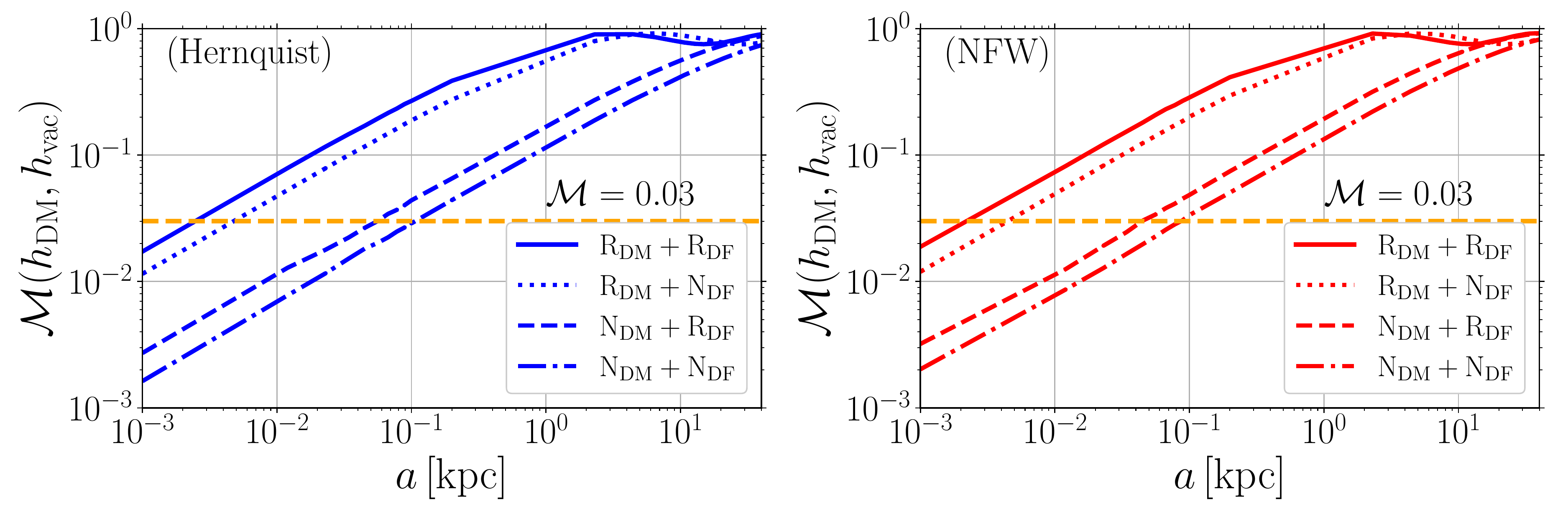}
\caption{
Top panels: mismatch plots as a function of the initial DM scale $\rho_0$, using the same binary configuration of Fig.~\ref{fig:EMRI_dephasing_Hernquist} for the vacuum waveform.  
Bottom panels:  mismatch plots as a function of the scale factor $a$. All of the configurations considered overcome the threshold $\mathcal{M}=0.03$ (marked by a dashed horizontal line) above which we expect the environmental effect to be detectable in a wide range of scale parameters.}
\label{fig:mismatches}
\end{figure*}

The number of GW cycles $\mathcal{N}_{\text{cycles}}$ is important for the detectability of GW signals. The change in the gravitational potential from DM, as well as the DF drag it induces,  will change the number of gravitational wave cycles during the inspiral process. This allows us to gain useful information about the DM through GW measurements. As a rule of thumb, any modification leading to a change in cycles $\Delta\mathcal{N}_\text{cycles}>1$ is potentially detectable. 

We start by asking some relevant questions. Are relativistic corrections to DF and the gravitational potential important? How do they compare to each other?  Do they lead to $\Delta\mathcal{N}_\text{cycles}>1$?
We answer these questions by estimating the number of cycles for three relativistic prescriptions for the DM spike and DF. One model, denoted by  {\quotes{$\text{R}_\text{DM}+\text{R}_\text{DF}$}}, includes relativistic corrections in both the DM spike and the DF [i.e., we set $\zeta = 1$ in $\xi(v)$ from Eq.~\eqref{DF_drag_force}]. In a second model, denoted by  {\quotes{$\text{R}_\text{DM}+\text{N}_\text{DF}$}}, we turn off the relativistic corrections to the DF (i.e., we set $\zeta=0$). In the third and final model,  {\quotes{$\text{N}_\text{DM}+\text{R}_\text{DF}$}}, we turn off relativistic corrections in the spike (by employing the NFW Newtonian model from Table~\ref{tab:fit_parameters}), but not in the DF. In all cases, we fix masses $10^5 M_\odot < M_\text{BH} < 10^7 M_\odot$, so that the signal is in the LISA band, and we consider initially NFW profiles with $\rho_0 = 0.3$GeV/cm$^3$ and $a=20$kpc. We integrate the expression for $\mathcal{N}_\text{cycles}$ in Eq.~\eqref{eq:Ncycles_def} from a starting frequency $f$ to the ISCO for all three prescriptions. We then calculate the difference $\Delta\mathcal{N}_\text{cycles}$ with a model without any relativistic corrections ({\quotes{$\text{N}_\text{DM}+\text{N}_\text{DF}$}}).

The results shown in Fig.~\ref{fig:cycles} allow us to estimate which relativistic corrections are most important. The pattern emerging from the figure is that it is crucial to include relativistic corrections to both the DF and to the DM spike.
The relativistic correction to the DM spike always induces larger dephasings than the relativistic correction to the DF. Relativistic corrections to the DF become relatively more important as the primary's mass is decreased, but
relativistic corrections to both the spike profile and DF need to be included for $M_\text{BH}\lesssim 10^7 M_\odot$. Only for $M_\text{BH}=10^7 M_\odot$ does the relativistic contribution to the DF become negligible, being barely visible in the plot. While this calculation refers to NFW, the results are qualitatively similar for initially Hernquist profiles.

This calculation also allows us to estimate the critical value of $\rho_0$ needed for a detection. A simple estimate can be obtained by numerically finding the value of $\rho_0$ which gives us $\Delta \mathcal{N}_{\text{cycles}}=1$ for several different observation times. The results are shown in Table~\ref{tab:crit_rho0}. As the BH mass $M_{\text{BH}}$ increases, the typical number of GW cycles in band decreases, and the local density values needed to have one cycle of dephasing become higher. As expected, we need a higher DM density for its effects to be observable if we consider shorter observation times.

What orders in the PN expansions are DM and DF contributions comparable to? To answer this question, we compare the contributions from DM-induced changes in the gravitational potential and from DF (both of which now include relativistic corrections) relative to the Newtonian, 1PN and 2PN contributions to the number of cycles of the binary in vacuum. The results are shown in Fig.~\ref{fig:dephasing} for a $(10^5+100)M_{\odot}$ binary and DM densities in the range $0.1< \rho_0[\text{GeV/cm}^3] < 0.5$. We see that DM and DF are always subdominant with respect to the Newtonian and 1PN contributions throughout the whole frequency range, but DF can dominate over 2PN terms for systems that are observed for roughly more than a year (see the crossing between the blue band and the 2PN terms, which roughly corresponds to the vertical dashed line). 
DM contributions to the gravitational potential are always subdominant to 2PN terms and to the DF drag.

These results depend on the total mass of the system. In Table~\ref{tab:table} we perform a more extensive comparison for a broader range of masses. As we decrease the total mass of the binary, the DF contributes at a level comparable to lower PN orders. This is consistent with the fact that the spike grows for lower masses, as seen in the scaling law of Eq.~\eqref{spike_scalings}.
Note also that the relativistic corrections always change the number of cycles in an appreciable way (especially for the DF), as shown in Fig.~\ref{fig:cycles}.  Increasing the mass to $10^6 M_\odot$ or even $10^7 M_\odot$, DF becomes subdominant to the 2PN order as well, but with an absolute number of cycles ($\sim 60$) that is well within the realm of detectability.

We can also compute the time one would need to observe the system so that DM+DF effects would begin dominating the 2PN contributions as a function of the binary's mass ratio. We vary the starting frequency $f_{\text{ini}}$ (which corresponds to some time before merger), and track the system all the way to the ISCO, using both the relativistic DM spike and the relativistic DF prescription. The necessary observing time is then found by separating out the different effects on the cycling, and numerically solving the equation $\mathcal{N}_{\text{DM+DF}}(f_{\text{ini}})=\mathcal{N}_{\text{2PN}}(f_{\text{ini}})$ for the necessary starting frequency $f_{\text{ini}}$.
The results of this calculation are shown in Fig.~\ref{fig:PN_comp_to_DM}: 2PN crossings with the DM and DF effects for reasonable observation timescales are only possible when $M_{\text{BH}}=10^5 M_\odot$. For higher primary mass systems, one would need $>10$~yr of observation. We also looked at the observation times that would be necessary for DM+DF to dominate 1PN contributions, and found that they always exceed $10$~yr. 

As the BH mass decreases, 2PN contributions tend to become subdominant for even lower observation timescales. This is because the spike grows for lower BH masses, as shown in Eq.~\eqref{spike_scalings}, while the number of cycles contributed by the 2PN correction shrinks. The larger spike has a larger impact on the number of cycles, and therefore we need less time for the DM+DF contributions to overtake the 2PN contributions. While we may expect the necessary observation time to continue decreasing for lower choices of primary mass, the halo feedback effect becomes important at such low masses~\cite{Kavanagh:2020cfn}. Halo feedback can deplete the spike, and hence decrease the dephasing due to DM.

The final question we address is: how are EMRIs affected by the inclusion of relativistic effects in the DM spike and DF?
We perform a dephasing and mismatch analysis using the EMRI models described in Sec.~\ref{sec:FEW_traj}.
As a first step, we obtain dephasings by tracking the change in polar phase between the FEW models with DM and in vacuum, $\Delta\Psi$, for each DM spike model in Table~\ref{tab:fit_parameters}.
The results are reported in Fig.\ref{fig:EMRI_dephasing_Hernquist}. All panels refer to a fully relativistic trajectory for an EMRI of primary mass $10^6M_\odot$ and secondary $10 M_\odot$ at a distance of $D_L=1$Gpc. The signals are observed for 4~years from a separation of $r = 15M_\text{BH}$, which is within the range in which we would expect differences between Newtonian and relativistic spikes. The optimal SNR for the vacuum waveform $h_{\text{vac}}$ for this EMRI configuration is $\rho \sim$46. Here we have defined the SNR $\rho \equiv \sqrt{(h_{\text{vac}}| h_{\text{vac}})}$ (not to be confused with the DM density), where $(\cdot|\cdot)$ is the inner product for real valued time-series
\begin{equation}\label{eq:inn_prod}
(a|b) = 4\text{Re}\int_{0}^{\infty}\frac{\hat{a}(f)\hat{b}^{\star}(f)}{S_{n}(f)}df,
\end{equation}
and $S_n(f)$ is the static power spectral density (PSD) for the LISA detector~\cite{Robson_2019}.
We have checked that the SNR is not affected appreciably by the DM effects.

In the top panel of Fig.~\ref{fig:EMRI_dephasing_Hernquist} we consider Hernquist profiles with and without relativistic corrections to the DM spike, and we use a Newtonian DF force to isolate the corrections coming from the spike. We vary the initial DM density around a nominal value $\rho_0 = 0.3$GeV/cm$^3$. Relativistic corrections to the DM spike play an important role in terms of detectability: Newtonian spikes lead to detectability (i.e., $\Delta\Psi\gtrsim 1$) after 3 years with an SNR of 38, while adding relativistic corrections would allow us to detect DM effects much earlier (after roughly 2 years in this example).
In the middle panel we repeat the analysis for NFW profiles, confirming these findings. 
Finally, in the lower panel we consider the relativistic corrections in the DF force discussed below Eq.~\eqref{DF_drag_force}. In the configuration we have chosen, the orbital velocity $v/c\sim 0.26$, and already in this case models of DF must include relativistic corrections. These results indicate that the effects of DF are large enough to be detected, confirming our earlier findings based on PN estimates of the number of cycles.

While the dephasing analysis is a good guide to understand which models are more likely to lead to detectable DM effects, it does not convey any information about the strength of the signal in the LISA frequency band. To include this information, we also perform a mismatch analysis. The mismatch is defined as 
\begin{equation}
    \mathcal{M} = 1 - \frac{(h_{\text{DM}}| h_{\text{vac}})}{\sqrt{(h_{\text{DM}}|h_{\text{DM}})(h_{\text{vac}}|h_{\text{vac}})}},
\end{equation}
where $h_{\text{DM}}$ and $h_{\text{vac}}$ are the waveforms generated with the DM-modified trajectory and the one in vacuum, respectively. A value of the mismatch between waveforms of $\mathcal{M}=0$ implies that the signals are identical, while values $\mathcal{M}\gtrsim 0.03$ imply that the mismatch could lead to observable effects \cite{Sathyaprakash:1991mt}.
In Fig.~\ref{fig:mismatches} we plot mismatches as a function of both $\rho_0$ and $a$. 
For plausible initial values of the DM spike overdensities, the mismatches are large enough to imply that DM effects can be detectable with EMRIs.
The oscillations of models with high mismatches are a result of an interference pattern between the waveforms with and without the spike as the number of dephasing cycles is increased: for a positive interference, the overlap between waveforms increases and the mismatch decreases. The resulting oscillations do not carry crucial physical information about the system, and one should be wary of comparing mismatches between models when they appear.

The mismatch calculation confirms our main conclusions from the calculation of dephasing cycles: relativistic corrections to the spikes are more important than relativistic corrections to DF, but both need to be included to avoid misinterpreting the inference of DM properties.

\section{Conclusions}\label{sec:conclusions}

The proposal to use IMRIs in the LISA band to assess the properties of DM overdensities is about a decade old~\cite{Gondolo:1999ef,Sadeghian:2013laa,Eda:2013gg,Eda:2014kra}. Recent work has carried out a more detailed detectability and measurability analysis, focusing mostly on Newtonian systems within a DM spike described by simple power-law models~\cite{Kavanagh:2020cfn,Coogan:2021uqv}.

In this paper we have highlighted the importance of including relativistic corrections to models of the DM spikes and (motivated by Refs.~\cite{Traykova:2021dua,Vicente:2022ivh}) in the expressions for the DF drag force. 
We have built catalogs for relativistic DM spike models, following the work of Sadeghian et al.~\cite{Sadeghian:2013laa}, but extending it to a much larger range of DM scale parameters $a$ and $\rho_0$. On the DF side, we have added prescriptions for the relativistic corrections to the classic Chandrasekhar DF formula~\cite{Chandrasekhar:1943ys} taken from Ref.~\cite{Barausse:2007ph,Traykova:2021dua}, but being careful not to overfit the model with a mixture of accretion and DM information~\cite{Vicente:2022ivh}.

We have assessed the importance of relativistic effects with a dephasing and mismatch study involving both a PN prescription for signals in circular orbits, and fully relativistic trajectories based on the Teukolsky equation for quasicircular orbits in a Schwarzschild background~\cite{Chua:2018woh,Chua:2020stf,Katz:2021yft,few}. These are some of our main findings:
\begin{itemize}
\item PN estimates of the number of cycles for a wide range of masses $10^5 M_\odot < M_\text{BH} < 10^7 M_\odot$ imply that relativistic corrections to the gravitational potential of the DM spike and to DF have  a considerable effect (see Fig.~\ref{fig:cycles}), and must be included in any realistic phenomenological model. 
\item The drag from DF dominates the 2PN contribution to the number of cycles for masses $M_\text{BH}\sim \mathcal{O}(10^5) M_\odot$ and observation times of $\gtrsim$ 1 year, but is subdominant for higher masses: see Fig.~\ref{fig:dephasing}, Fig.~\ref{fig:PN_comp_to_DM}, and Table~\ref{tab:table}. 

\item  The inclusion of torques from DF in fully relativistic FEW models leads to large dephasings and mismatches for a selected sample of EMRI configurations (see Figs. \ref{fig:EMRI_dephasing_Hernquist} and \ref{fig:mismatches}). This suggests that DF is an important contribution that should be accounted for in EMRI data analysis, if such systems are formed within DM spikes.
\end{itemize}

The DM models used here are easily included within the FEW framework, which is already capable of rapid waveform generation. The additional torques do not jeopardize the speed of waveform generation, and therefore environmental effects such as those from DM spikes can be included in current EMRI/IMRI models for LISA data analysis without significant computational overhead.

Our work lends itself to several follow-ups. First of all, it would be interesting to carry out a measurability study to constrain the precision with which the DM parameters can be estimated in LISA data analysis.
Furthermore, it would be interesting to consider other prescriptions for either the underlying nature of DM (see the fuzzy-DM scenarios~\cite{Hui:2016ltb}) and how they relate to specific DF prescriptions~\cite{Vicente:2022ivh}. From the theoretical point of view, it is very important to study EMRI dynamics beyond the circular-orbit limit considered here (see e.g. Ref.~\cite{Cardoso:2020iji}).
From the astrophysical point of view, much work is needed to estimate the rates of EMRI/IMRI events that may be affected by DM spikes, or to understand whether environmental effects due to DM can be disentangled from (say) the presence of an accretion disk~\cite{Kocsis:2011dr}. We leave all of these topics to future work.

\acknowledgments

The authors wish to thank Lorenzo Speri and Andrea Caputo for very fruitful discussions.
This work makes use of the Black Hole Perturbation Toolkit~\cite{BHPToolkit}.
N.S., A.A. and E.B. are supported by NSF Grants No. PHY-1912550, AST-2006538, PHY-090003 and PHY-20043, and NASA Grants No. 17-ATP17-0225, 19-ATP19-0051 and 20-LPS20-0011. This research project was conducted using computational resources at the Maryland Advanced Research Computing Center (MARCC). 
The following software libraries were used at various stages in the analysis for this work, in addition to the packages explicitly mentioned above: \texttt{numpy} \citep{harris2020array}, \texttt{matplotlib} \citep{Hunter:2007}, \software{scipy} \citep{2020SciPy-NMeth}, \software{Mathematica} \citep{Mathematica}, \software{filltex}~\cite{2017JOSS....2..222G}.

\bibliography{refs}

\end{document}